\documentclass[%
    reprint,
    superscriptaddress,
    nofootinbib,
    amsmath,amssymb,
    aps,
    ]{revtex4-2}

\usepackage{hyperref}
\usepackage{cleveref,csquotes,placeins}
\usepackage[acronym,nolong]{glossaries}
\usepackage{lipsum}
\usepackage{graphicx}
\usepackage{dcolumn}
\usepackage{bm}
\usepackage{orcidlink}
\usepackage{booktabs}
\usepackage{mathtools,mathrsfs}
\usepackage{bbm}
\usepackage[shortlabels]{enumitem}
\usepackage{comment}

\DeclarePairedDelimiterX\braket[2]{\langle}{\rangle}{#1\,\delimsize\vert\,\mathopen{}#2}

\usepackage{pgfplots}
\pgfplotsset{compat=newest}
\usepgfplotslibrary{external}
\usetikzlibrary{spy}
\usepgfplotslibrary{groupplots}
\pgfdeclarelayer{background}
\pgfdeclarelayer{foreground}
\pgfsetlayers{background,main,foreground}
\usepackage{subcaption}

\glsdisablehyper
\begin{document}

\title{Split Ends: UV Physics branches Black Hole Hair}

    \author{Georg Trenkler\,\orcidlink{0009-0009-5206-6865}}
	\email[]{gtrenkler@ibs.re.kr}
    \affiliation{Cosmology, Gravity, and Astroparticle Physics Group, Center for Theoretical Physics of the Universe, Institute for Basic Science (IBS), Daejeon, 34126, Korea.}
 
	\author{Leonardo G. Trombetta\,\orcidlink{0000-0001-7345-5100}}
	\email[]{leonardo.trombetta@ib.edu.ar}
    \affiliation{Centro At\'omico Bariloche and Instituto Balseiro,\\ 
    S.C. de Bariloche, R\'io Negro, R8402AGP, Argentina}

    \author{Alexander Vikman\,\orcidlink{0000-0003-3957-2068}}
	\email[]{vikman@fzu.cz}
	\affiliation{CEICO, FZU-Institute of Physics of the Czech Academy of Sciences, Na Slovance 1999/2, 182 00, Prague 8, Czech Republic}

\begin{abstract}
    Coupling a shift-symmetric scalar field to the Gauss-Bonnet invariant modifies black holes (BHs) via scalar hair. Here we consider this field to be the Goldstone of a complex scalar field with a symmetry-breaking potential. For BHs of a size comparable with the inverse mass gap, the gapped field tries to restore the symmetry close to the horizon, enhancing the departure from standard BHs. Moreover, BHs of the same mass can have up to three different horizon areas and surface gravities. This branching has profound consequences for BH physics, potentially observable in gravitational waves.
\end{abstract}

\maketitle

{ \textbf{\textit{Introduction}} }
Symmetries provide a fundamental guiding principle for constructing
low-energy effective field theories (EFTs). In particular, the Goldstone
theorem ensures that spontaneously broken continuous global symmetries
result in massless collective degrees of freedom -- Nambu-Goldstone
bosons. These bosons acquire a parametrically small mass if the symmetry
is approximate. Light bosonic degrees of freedom are well suited for
modeling large-scale phenomena such as dark matter (DM), dark energy
(DE) and infrared (IR) modifications of gravity. From the EFT perspective, these light degrees of freedom generically couple to spacetime curvature, thereby effectively modifying gravity. In four spacetime dimensions the lowest
order curvature invariants are: parity-even $R$, $R_{\mu\nu}R^{\mu\nu}$,
$R_{\alpha\beta\mu\nu}R^{\alpha\beta\mu\nu}$ and the parity-odd Pontryagin
term $\mathcal{P}=R_{\alpha\beta\mu\nu}\,^{*}R^{\alpha\beta\mu\nu}$. The simplest Nambu–Goldstone boson, a massless real scalar field $\phi$, enjoys a global shift symmetry $\phi\rightarrow\phi+c$, with constant $c$. This symmetry is preserved by linear couplings to topological terms like parity-odd 
$F_{\mu\nu}\,^{*}F^{\mu\nu}$ for a vector gauge field and $\mathcal{P}$ for gravity, or to the parity-even Gauss-Bonnet term $\mathcal{G} = - R_{\alpha\beta\mu\nu}\,^{*}R^{*\alpha\beta\mu\nu}$
in four spacetime dimensions. Here 
\begin{eqnarray}
& ^{*}F^{\mu\nu}=\frac{1}{2}\varepsilon^{\mu\nu\alpha\beta}F_{\alpha\beta}, & \\
& ^{*}R^{\mu\nu\rho\sigma}=\frac{1}{2}\varepsilon^{\mu\nu\alpha\beta} R_{\alpha\beta}{}^{\rho\sigma}\,,\quad\,^{*}R^{*\mu\nu\rho\sigma}=\frac{1}{2}\varepsilon^{\alpha\beta\rho\sigma}\,^{*}R^{\mu\nu}{}_{\alpha\beta}\,, & \notag
\end{eqnarray}
where $\varepsilon^{\mu\nu\alpha\beta}$ is the Levi\nobreakdash-Civita tensor. Note that
shift-preserving couplings to other invariants necessarily involve more
powers of $\phi$ along with its derivatives. Thus, these operators
have higher mass dimensions and are more irrelevant in the IR. Moreover,
due to the difference in parity the couplings to $\mathcal{P}$ and to $\mathcal{G}$ can be parametrically different. In addition, the linear coupling to $\mathcal{G}$ keeps the
equations of motion second order\footnote{Such couplings belong to the
Horndeski class~\cite{Horndeski:1974wa}, as shown in~\cite{Kobayashi:2011nu}.}, whereas a linear coupling to $\mathcal{P}$ leads to higher-derivative equations of motion. Thus, a coupling to $\mathcal{G}$ can be studied not only perturbatively but also in the fully nonlinear regime. Moreover, it is technically natural to assume that even an approximate shift symmetry suppresses couplings to other non-topological curvature invariants by a parametrically higher scale. 
High-curvature environments, like BHs, offer a natural laboratory for probing such EFTs. In particular, theories with a linear coupling to the Gauss--Bonnet term (sGB) endow BHs with
a nontrivial scalar field configuration---scalar hair~\cite{Sotiriou:2013qea,Sotiriou:2014pfa}---in the form of a scalar charge, thereby evading the no-hair theorem for shift-symmetric scalars~\cite{Hui:2012qt, Creminelli:2020lxn}. Such charges can have important phenomenological consequences for BH binaries, most notably through scalar radiation, which accelerates the inspiral, and can even generate dipolar radiation in compact binaries~\cite{Yagi:2011xp}, providing a sensitive observational probe~\cite{Lyu:2022gdr,Sanger:2024axs}. Additional shift-symmetric interactions can modify BH charges and conditions for existence of the static hair~\cite{Noller:2019chl,Thaalba:2022bnt,Thaalba:2025lwe,Lara:2025hqh}. One can also ignore the natural advantage of shift symmetry and include nonlinear couplings to $\mathcal{G}$ term to study BHs \cite{Callan:1988hs,Mignemi:1992nt,Kanti:1995vq,Kokkotas:2017ymc,Doneva:2017bvd,Silva:2017uqg,Andreou:2019ikc}. Interestingly, some of these couplings appear in string theory \cite{Zwiebach:1985uq,Gross:1986mw,Metsaev:1987zx,Antoniadis:1992rq}. 
Cosmological tachyonic instabilities constrain~\cite{Anson:2019uto} some popular theories with nonlinear couplings to $\mathcal{G}$. 

In this work we consider a minimal UV extension of linearly coupled sGB where the Goldstone arises from the spontaneously broken $U(1)$-symmetry of a complex scalar field. This is unlike other existing constructions that do not preserve shift symmetry, and e.g. produce scalar walls through curvature-induced symmetry restoration~\cite{Babichev:2022djd}, hairy BHs through curvature-induced $U(1)$ breaking~\cite{Latosh:2023cxm,JyotiDas:2026qlq}, or couple linearly to the QCD axion~\cite{Karananas:2026nok}. Note that a complex scalar is an interesting light DM candidate, see \cite{Li:2013nal,Li:2016mmc,Fan:2016rda,Suarez:2017mav}. For our symmetry-breaking potential, integrating out the gapped mode--the modulus, $\rho$,--at leading derivative order, see e.g.~\cite{Babichev:2018twg,Glavan:2025khe}, yields derivative interactions consistent with flat-space positivity bounds~\cite{Adams:2006sv}, contrary to the screening and self-accelerating models of Ref.~\cite{Lara:2025hqh}. Our construction provides an incomplete UV-extension of sGB, thus not addressing issues like asymptotic causality \cite{Serra:2022pzl, Bucciotti:2026svg}. Around BHs the geometrically sourced Goldstone hair pushes $\rho$ away from its vacuum expectation value. This $\rho$-\emph{activation} breaks the effective single-field description in the matter sector. This can be also relevant for other non-static configurations around BHs e.g.~\cite{Hui:2019aqm,Feng:2021qkj,Glavan:2025khe}. Here we calculate static spherically-symmetric BH solutions in this theory perturbatively and numerically. Our results show that smaller BHs exhibit stronger $\rho$-\emph{activation} and larger departures from GR compared to sGB, across several BH observables. The impact of the new physics becomes maximal for sufficiently small BHs, where a \emph{novel} multi-branch behavior emerges as seen from numerical solutions. We briefly discuss various phenomenological implications and formal aspects of BH physics that need revisiting in this context.


{ \textbf{\textit{Dynamics and Scales}} }
It is convenient to parametrize\footnote{We use $M_{\text{P}}=(8\pi G_{\text{N}})^{-1/2}$ and $\hbar=c=1$.} the complex field as $\Psi=M_{\text{P}}\,\rho\,e^{i\phi/\rho_{0}}$,
where $\rho_{0}$ corresponds to the global minimum of the $U(1)-$invariant
potential $V\left(\rho\right)$. Using dimensionless coordinates $x=y/\ell$,
where $\ell$ is a length scale of interest, we write the action of
the canonical $\Psi$, through which the Goldstone $\phi$ couples
to the rescaled Gauss-Bonnet invariant $\mathcal{G}$ via the dimensionless
coupling $\alpha$,
\begin{align}
 S\left[g,\rho,\phi\right] =& \int d^{4}x\sqrt{-g}\left[-\frac{1}{2}R+\frac{1}{2}\left(\partial\phi\right)^{2}+\alpha\,\phi\,\mathcal{G}\right.
 \notag\\
 & \left.+\frac{1}{2}\left(\partial\rho\right)^{2}-V\left(\rho\right)+\frac{\rho^{2}-\rho_{0}^{2}}{2\rho_{0}^{2}}\,\left(\partial\phi\right)^{2}\right]\,,\label{action-CsGB}
\end{align}
where we omitted the overall factor $\left(M_{\text{P}}\ell\right)^{2}$. The first line of this action corresponds to sGB. 
Many of our formulas are valid for a general potential $V(\rho)$,
but for concreteness we use the renormalizable potential
\begin{equation}
V\left(\rho\right)=-\frac{\mu^{2}}{2}\rho^{2}+\frac{\lambda}{4}\rho^{4}+\frac{\mu^{4}}{4\lambda}\,,
\label{potential}
\end{equation}
so that that the vacuum is given by $\rho_{0}=\mu/\sqrt{\lambda}$, and $V(\rho_0)=0$. Thus, the theory is fully characterized by three dimensionless parameters
$\mu, \lambda, \alpha$ and a length scale $\ell$. The canonical mass parameter and
coupling constant are given by $\mu_{\text{ph}}=\mu/\ell$ and $\lambda_{\text{ph}}=\lambda/\left(\ell M_{\text{P}}\right)^{2}$
respectively. Fixing $\rho=\rho_{0}$, the irrelevant interaction of the canonically normalized Goldstone to physical $\mathcal{G}$ is suppressed\footnote{The corresponding strong-coupling scale $(M_{\text{P}}/\alpha \ell^2)^{1/3}$.} by the mass scale $\left(\alpha\ell^{2}M_{\text{P}}\right)^{-1}$. On the other hand, in geometrical units often adopted in the literature, e.g. Refs. \cite{Sotiriou:2013qea, Sotiriou:2014pfa}, we identify $\alpha_{\text{g}}=\alpha\ell^{2}$. No nonminimal couplings of $\rho$ to curvature are assumed\footnote{These are radiatively generated with usual loop factor suppression, and negligible for small $\lambda_{\text{ph}}$.}

The equations of motion for the scalar components are 
\begin{eqnarray}
    \square \rho - \frac{\rho}{\rho_0^2} (\partial \phi)^2 + V_{,\rho}(\rho) &=& 0, \label{EOM-rho} \\
    \nabla_\mu \left( \frac{\rho^2}{\rho_0^2} \nabla^\mu \phi \right) &=& \alpha \mathcal{G}\;, \label{EOM-phi}
\end{eqnarray}
with $V_{,\rho}\equiv\frac{\partial V}{\partial\rho}$, and the Einstein equation is $G_{\mu\nu} = T_{\mu\nu}$, with $G_{\mu\nu}$ the Einstein tensor, and
\begin{eqnarray} \label{Tmunu-CsGB}
 T_{\mu\nu} &=& \partial_\mu \rho \partial_\nu \rho + \frac{\rho^2}{\rho_0^2} \partial_\mu \phi \partial_\nu \phi \\
 &&- g_{\mu\nu} \left( \frac{1}{2} (\partial \rho)^2 + \frac{1}{2} \frac{\rho^2}{\rho_0^2} (\partial \phi)^2 - V(\rho) \right) \notag \\
 &&- 2 \alpha \, g_{\rho (\mu} g_{\nu) \delta} \, \nabla_{\sigma} \left( \partial_{\gamma} \phi \, \epsilon^{\gamma\delta\alpha\beta} \epsilon^{\rho\sigma\lambda\eta} R_{\lambda\eta\alpha\beta} \right) , \notag
\end{eqnarray}
the Stress-Energy tensor.

{ \textbf{\textit{Static, Spherically-Symmetric BHs}} }
We now seek static, asymptotically flat BH solutions
\begin{equation}
    \begin{gathered}
        ds^2=h(r)\,dt^2-\dfrac{dr^2}{f(r)}-r^2d\Omega^2,\\
        \rho=\rho(r)\,,\qquad \phi=\phi(r)\,.
    \end{gathered}
    \label{eq:ansatz}
\end{equation}
In the following, radial derivatives and evaluation at the horizon are denoted by $'$ and subscript $h$ respectively.
Because the phase is shift-symmetric, Eq.~\eqref{EOM-phi} is a radial
conservation law and admits the exact first integral
\begin{equation}
    -\frac{\rho^2}{\rho_0^2}\phi'
    +\frac{4\alpha h'}{r^2h}(1-f)
    =\frac{Q}{r^2\sqrt{hf}}\,,
    \label{phase-eq-int-once}
\end{equation}
where $Q=const$, which for an asymptotically flat solution with $\rho\to\rho_0$ is precisely the scalar charge associated to the phase field, since
$\phi=\phi_\infty+Q/r+\mathcal{O}(r^{-2})$.

Assuming $\rho_h$ and $\phi'_h$ are finite near a nonextremal horizon $f(r_h)=h(r_h)=0$,
matching the simple poles in Eq.~\eqref{phase-eq-int-once} yields the exact relation
\begin{equation}
    Q=8\alpha\kappa\,,
    \label{eq:charge}
\end{equation}
where $\kappa = \frac{1}{2} \lim_{r \to r_h} \sqrt{\frac{f}{h}} h'$ is the surface gravity, still obeying the charge--surface-gravity relation~\cite{Prabhu:2018aun}. Thus, as in standard sGB, the scalar charge is secondary: it is fixed by the horizon geometry rather than supplied as
independent data. The modulus $\rho$ affects $Q$ only indirectly, through its
backreaction on $\kappa$.

A near horizon expansion of the field equations determines the radial data
algebraically. In particular,
\begin{equation}
    f_h'=\frac{1-r_h^2 V(\rho_h)}{r_h+4\alpha\phi'_h},
    \qquad
    \rho'_h=\frac{V_{,\rho}(\rho_h)}{f_h'}\,,
    \label{eq:horizon-data}
\end{equation}
while $\phi'_h$ obeys a quadratic equation, with one root recovering the known sGB solutions in the proper limit.  
For given $\mu, \lambda, \alpha$ and the chosen root, the pair $(r_h,\rho_h)$ fixes $\phi'_h$, $f'_h$, and $\rho'_h$, while $h'_h$ and $\phi_h$ remain free until the time normalization at infinity and $\phi_\infty$ are fixed. Reality of the solution for $\phi'_h$ requires
\begin{eqnarray}\label{existence}
    \mathcal{D}_h &\equiv{}& \frac{\rho_h^4}{\rho_0^4}
    -192\frac{\alpha^2}{r_h^4} \frac{\rho_h^2}{\rho_0^2}
    \\
    &&+128 
    \frac{\alpha^2 V_h}{r_h^2}
    \left(\frac{\rho_h^2}{\rho_0^2}-48\frac{\alpha^2}{r_h^4} + 8\frac{\alpha^2 V_h}{r_h^2} \right) >0\,. \notag    
\end{eqnarray}
This local condition must be supplemented by $1-r_h^2V_h>0$ and $r_h+4\alpha\phi'_h > 0$ for continuous sGB and GR limits, respectively. For $\rho_h=\rho_0$ and $V_h=0$, it reduces to the standard sGB minimum-size bound $r_h^4>192\alpha^2$ \cite{Sotiriou:2014pfa}.


{ \textbf{\textit{Perturbative Example}} }
Having established the regularity conditions that solutions must satisfy at $r=r_h$, we now turn to solving the full system of field equations, for potential \eqref{potential}. We begin by considering a perturbative expansion in $\alpha/r_h^2$ of Eqs.~\eqref{EOM-rho}, \eqref{EOM-phi}, and \eqref{Tmunu-CsGB}.
However, this approximation is insufficient to solve for $\rho$, which satisfies an effective equation obtained by replacing the solution for $\phi'$ into Eq.~\eqref{EOM-rho}. In terms of $u \equiv \rho/\rho_0$:
\begin{equation}   
\frac{1}{\mu^2r^2} \sqrt{\frac{f}{h}} \left( r^2 \sqrt{hf} u' \right)' - \frac{16 \ell_\text{NL}^6}{u^3} S(r) + u - u^3 = 0\,,
\label{eq:effective_rho_eq}
\end{equation}
where the two relevant dimensionless parameters are $\mu^{-1}$ and $\ell_\text{NL} = (\alpha \sqrt{\lambda}/\mu^2)^{1/3}$ --- length scales in units of $\ell$ --- while $S(r) = (f/r^4)[h'(1-f)/h - 2\kappa/\sqrt{hf}]^2$ acts as a geometrical source term.

Standard sGB is recovered in the heavy modulus limit $\mu \to \infty$, freezing it at $\rho = \rho_0$. Small departures from sGB are parametrized by $\ell_\text{NL}^6/
r_h^6$ with generally small coefficients. 

To proceed analytically, we consider instead the more interesting regime with $\mu^{-1} \gg r_h$, where the field $\rho$ is very light. At leading order we may drop any potential terms, and the gradient term is then sourced proportionally to  $\ell_\rho^4$, with the effective length scale 
\begin{equation}
\label{eq:ELL_rho}
\ell_\rho \equiv  \left(\mu^2 \ell_\text{NL}^6 \right)^{1/4}=\left(\frac{\lambda\alpha^{2}}{\mu^{2}}\right)^{1/4}\,.
\end{equation} 
Therefore, the horizon value $\rho_h$ is found to be
\begin{eqnarray} \label{eq:rho_h-pert}
    u_h = \frac{\rho_h}{\rho_0} &\simeq& 1 - 16 c_\rho \frac{\ell_\rho^4}{r_h^4}\,.
\end{eqnarray}
where, $c_\rho = \frac{3511}{3600} \simeq 0.97$. As expected, the sourcing of the $\phi$ field by the strong curvature near the horizon, and its coupling with $\rho$, causes the latter to depart from its minimum toward symmetry restoration.

We may go on to compute various quantities of interest associated with the BH. On an asymptotically flat solution, the mass and scalar charge can be read from an $r \to \infty$ expansion,
\begin{align}    
    f&=1-\frac{2M}{r}+\mathcal{O}(r^{-2})\,,\; h = 1-\frac{2M}{r}+\mathcal{O}(r^{-3})\,, \label{eq:asymptotics_h}\\
    \phi&=\frac{Q}{r}+\mathcal{O}(r^{-2})\,, \label{eq:asymptotics_phi} \\
    u &= 1- \frac{\lambda Q^2}{2 (\mu r)^4} +\mathcal{O}(r^{-5})\,. \label{eq:asymptotics_u}
\end{align}
Hence the massive modulus has no independent $1/r$ charge. The physical mass $M$ reads, 
\begin{eqnarray}
    M &\simeq& m + \frac{49 \, {\alpha}^{2}}{40 \, m^{3}} \Biggl( 1 + c_M \frac{\ell_\rho^4}{m^{4}} \Biggr)\,,
    \label{eq:M_perturbative}
\end{eqnarray}
where $m = r_h/2$ is the would-be BH mass in GR for fixed $r_h$, and $c_M = \frac{60171961}{34927200} \simeq 1.72$. The $\rho$-induced contribution to the energy is comparable with the sGB one when $\ell_\rho \sim m$. Other quantities are better expressed directly in terms of $M$, such as horizon area $A_h = 4\pi r_h^2$,
\begin{eqnarray}
    A_h &\simeq& 16 \, \pi M^{2} - \frac{196 \, \pi {\alpha}^{2}}{5 \, M^{2}} \left( 1 + c_A \frac{\ell_\rho^4}{M^4} \right) \, ,
    \label{eq:A_h_perturbative}
\end{eqnarray}
with $c_A = c_M$, horizon value of the phase field,
\begin{eqnarray}
    \phi_h \simeq \frac{11 \, {\alpha}}{6 \, M^{2}} \left( 1 + c_\phi \frac{\ell_\rho^4}{M^4} \right) \, ,
    \label{eq:phi_h_perturbative}
\end{eqnarray}
with $c_\phi = \frac{111017}{83160} \simeq 1.33$, and surface gravity,
\begin{eqnarray}
    \kappa &\simeq& \frac{1}{4 \, M} + \frac{73 \, {\alpha}^{2}}{240 \, M^{5}} \left( 1 + c_\kappa \frac{\ell_\rho^4}{M^4} \right) \, .
    \label{eq:kappa_perturbative}
\end{eqnarray}
with $c_\kappa = \frac{22423691}{13490400} \simeq 1.66$. The latter also allows to know the scalar charge through the exact relation \eqref{eq:charge}. These expressions agree with the standard sGB results at the LO in $\alpha$, for $\ell_\rho = 0$.

Interestingly, the new corrections to all the quantities above have the same signs as the sGB ones themselves. Therefore, in general we expect an enhanced departure from GR in this theory compared to sGB at a given fixed mass $M$, which is not simply a dressing of $\alpha$. 


{ \textbf{\textit{Numerical Solutions}} }
We now turn to a numerical study of static BH solutions to the full system of equations in spherical symmetry. The $rr$ Einstein equation is a quadratic algebraic constraint for $f^{-1}$. We choose the root satisfying $f\to1$ at infinity, differentiate the constraint once, and eliminate $f'$ from the $tt$, modulus, and phase field equations. 

We initialize the integration just outside the horizon using the regular
series, choose the appropriate root for $\phi'_h$, and use $\rho_h$ as the
shooting parameter. At $r \to \infty$ we impose $h\to1$, $\rho\to\rho_0$, and $\phi_\infty = 0$.

\begin{figure*}
\centering
\begin{subfigure}{0.49\textwidth}
    \includegraphics[width=\textwidth]{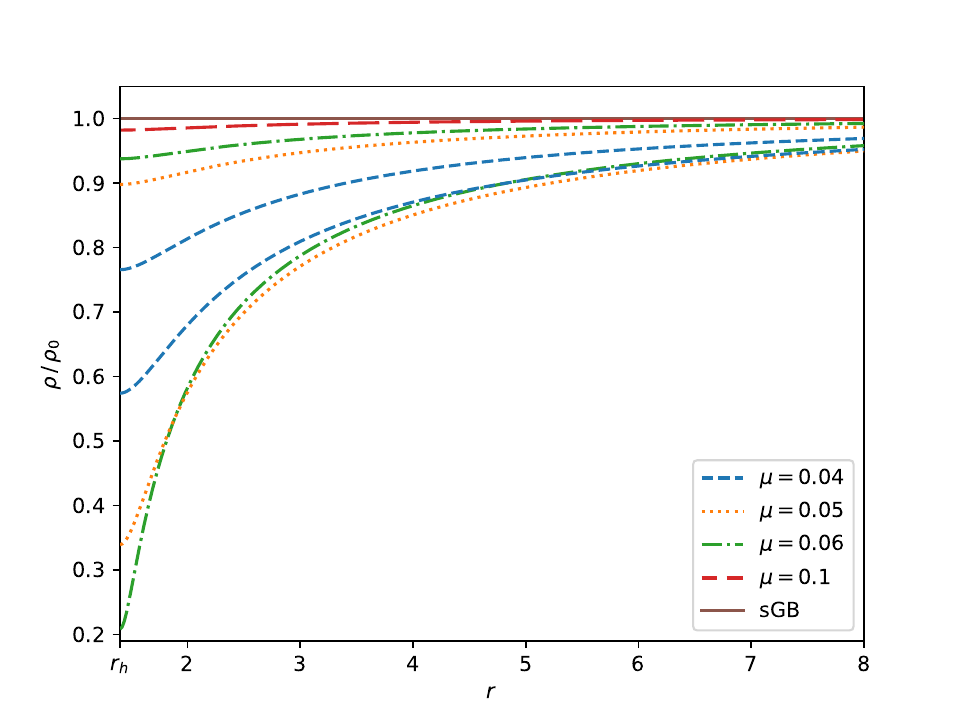}
    \caption{Profile of $u(r) = \rho(r)/\rho_0$}
    \label{fig:rho-vs-r}
\end{subfigure}
\hfill
\begin{subfigure}{0.49\textwidth}
    \includegraphics[width=\textwidth]{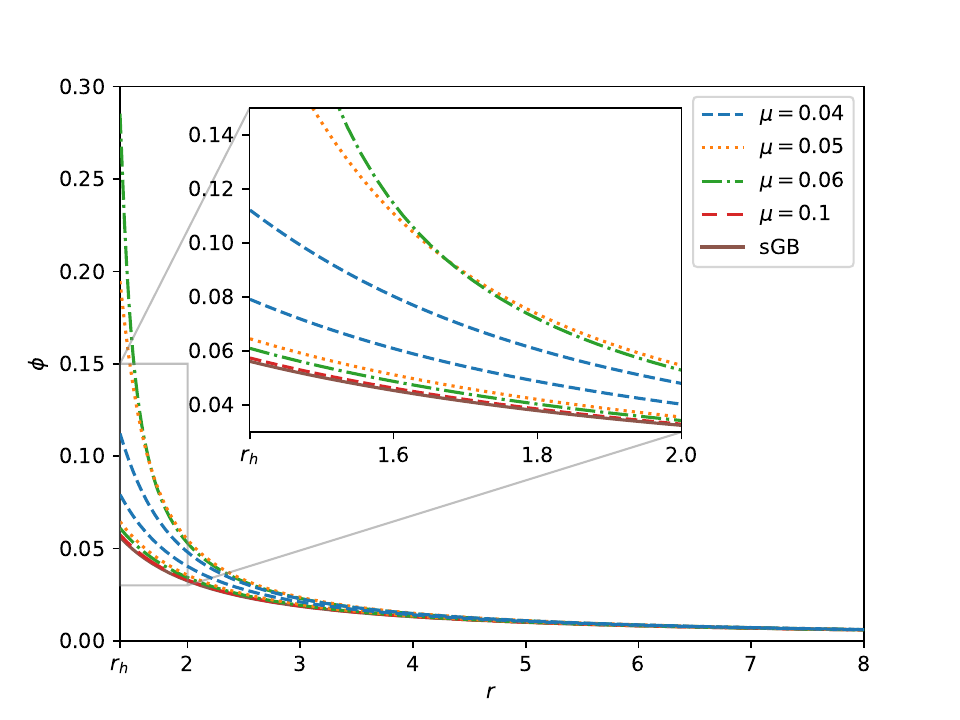}
    \caption{Profile of $\phi(r)$}
    \label{fig:phi-vs-r}
\end{subfigure}
\caption{Scalar hair of black-hole solutions with $r_h = 1.4$. Here $\alpha=0.015$ and $\lambda=0.25$}
\label{fig:r-dep}
\end{figure*}

In Fig.~\ref{fig:r-dep} we plot the profiles $\rho(r)$ and $\phi(r)$ for the fiducial value $r_h = 1.4$ and varying $\mu$. Interestingly, a \emph{novel} behavior emerges for some $\mu$ values: multiple solutions are observed to coexist at fixed $\mu, \lambda, \alpha$, and $r_h$ but different $\rho(r_h)$. The new solutions exhibit a smaller $\rho(r_h)$---toward symmetry restoration---(see Fig.~\ref{fig:rho-vs-r}), a larger $\phi(r_h)$ (see Fig.~\ref{fig:phi-vs-r}), and are inaccessible from the perturbative calculation above. We shall refer to them as \emph{active}-$\rho$ branches.

\begin{figure}[ht!]
    \includegraphics[width=0.49\textwidth]{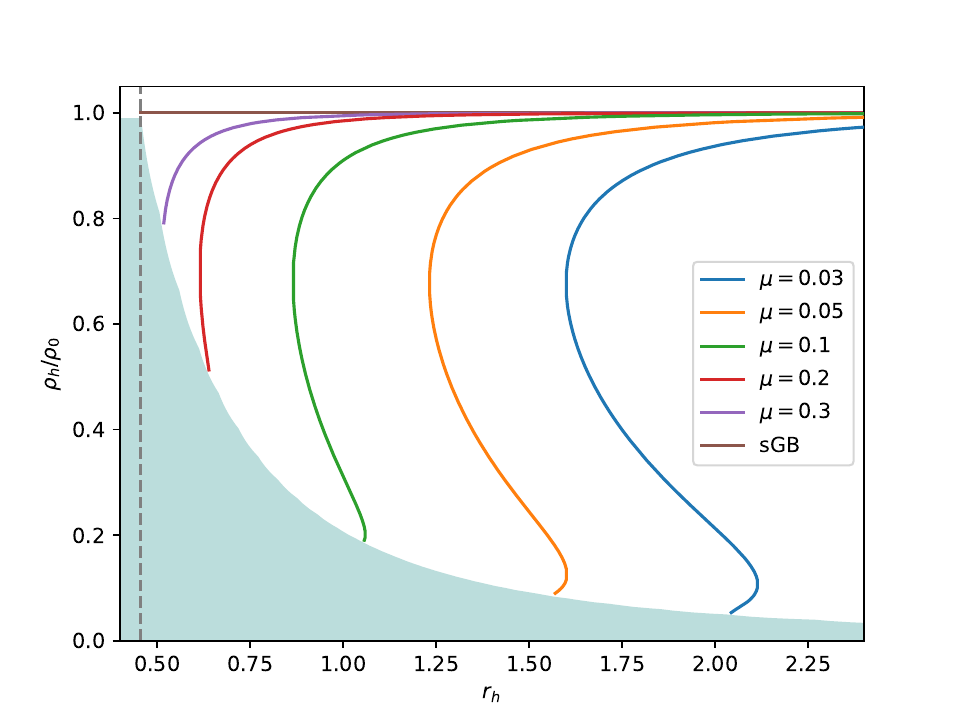}
    \caption{Modulus field at the horizon for $\alpha=0.015$ and $\lambda=0.25$. Each point on the curve is a BH solution of size $r_h$, with $\rho(r_h) = \rho_h$. The shaded region is excluded by condition \eqref{existence}. The dashed vertical line corresponds to the minimal possible $r_h$ in sGB.  
    }
    \label{fig:rhoh-vs-rh}
\end{figure}

We explore this branching behavior by fixing $\alpha=0.015$ and $\lambda=0.25$ and varying $\mu$, as we look for BH solutions of different sizes $r_h$. In Fig.~\eqref{fig:rhoh-vs-rh} we observe that larger BHs are closer to sGB, while smaller ones may experience a departure from sGB depending on $\mu$. Where solutions exist, up to three branches were found at fixed size $r_h$ for sufficiently small $\mu$. Moreover, there are no solutions below a minimum $r_h(\mu)$, which is always larger than the sGB value 
$(192\alpha^2)^{1/4}$ (gray dashed), and no longer generally determined by the condition $\mathcal{D}_h = 0$. A single sGB-like branch always persists at large $r_h$.

\begin{figure*}
\centering
\begin{subfigure}{0.49\textwidth}
    \includegraphics[width=\textwidth]{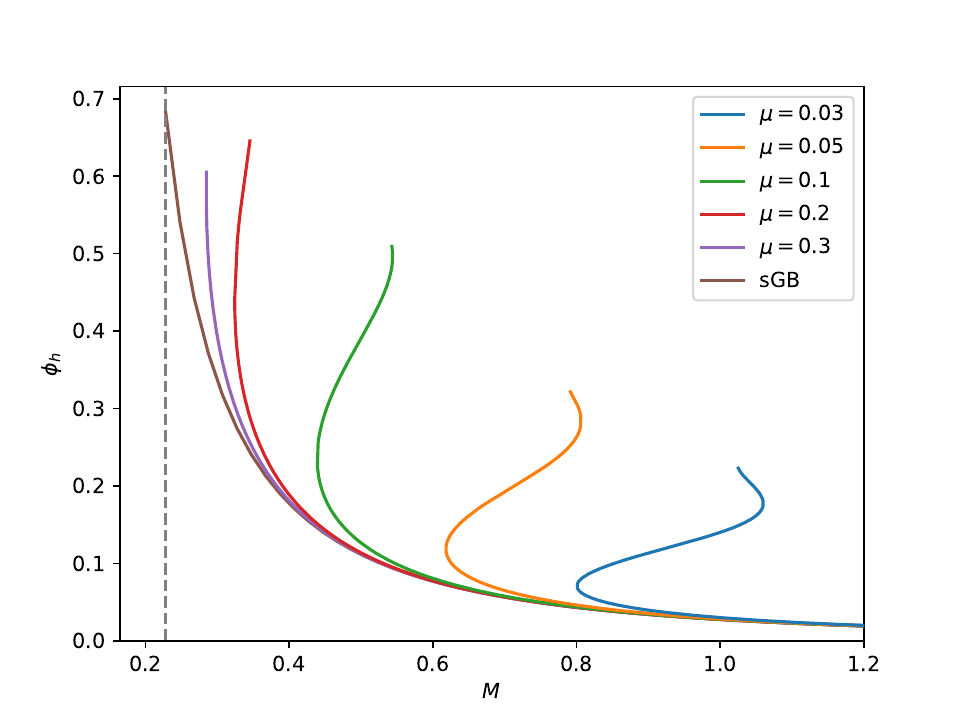}
    \caption{Phase field $\phi_h = \phi(r_h)$ at horizon.}
    \label{fig:phih-vs-M}
\end{subfigure}
\hfill
\begin{subfigure}{0.49\textwidth}
    \includegraphics[width=\textwidth]{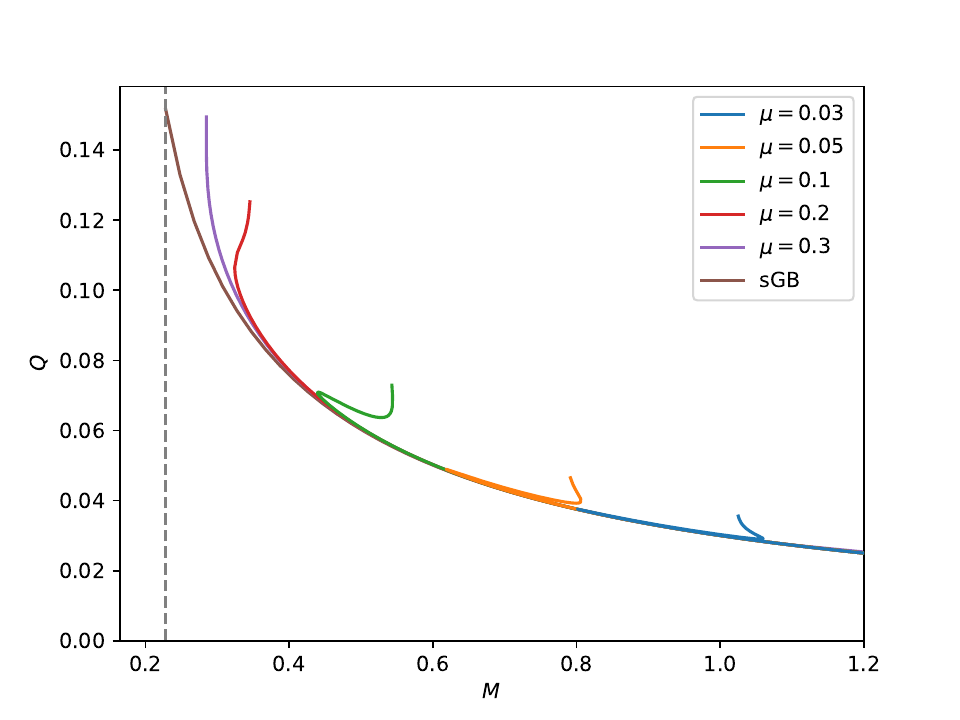}
    \caption{Scalar charge $Q$.}
    \label{fig:Q-vs-M}
\end{subfigure}
\hfill
\begin{subfigure}{0.49\textwidth}
    \includegraphics[width=\textwidth]{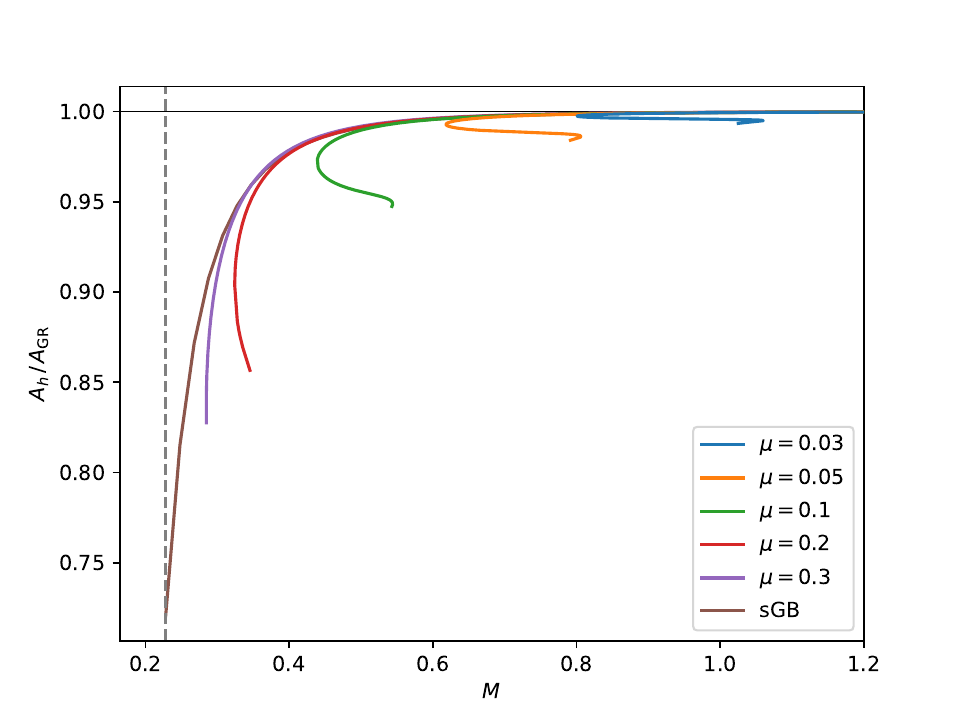}
    \caption{Horizon area $A_h$ relative to GR.}
    \label{fig:Ah-vs-M}
\end{subfigure}
\hfill
\begin{subfigure}{0.49\textwidth}
    \includegraphics[width=\textwidth]{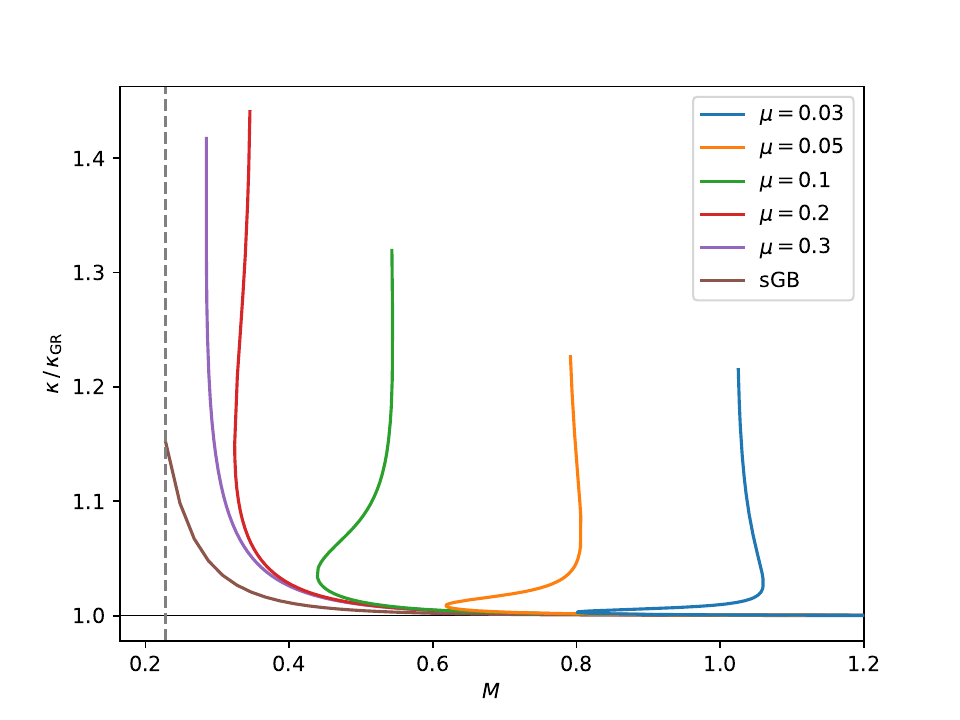}
    \caption{Surface gravity $\kappa$ relative to GR.}
    \label{fig:kappa-vs-M}
\end{subfigure}
\caption{Physical mass, $M$, dependence of BH parameters for $\alpha=0.015$ and $\lambda=0.25$. The dashed vertical line corresponds to the minimal possible $M$ in sGB. Generally, at large enough $M$ there is a unique solution that approaches the standard sGB one, coinciding with it when $\rho_h \to \rho_0$ (see Fig.~
\ref{fig:rho-vs-r}). As $M$ decreases below a $\mu$-dependent threshold, $\rho$ becomes active causing a number of effects: a) $\phi_h \sim 1/\rho_h^2$ increases (see Eq.~\eqref{phase-eq-int-once}), b) $Q$ increases, strengthening scalar-radiation bounds, and even producing monopolar radiation via branch transitions. Moreover, $dM/dQ\propto dM/d\kappa$ can change sign hinting to positive heat capacity. c) a larger fraction of the total energy is stored in the scalar configurations, decreasing $A_h$, and d) $\kappa$ increases against GR due to stronger backreaction. The latter two aspects can influence thermodynamics. When the \emph{active}-$\rho$ branches appear, they always have smaller $\rho_h$ for the same $M$, further enhancing all these effects. Some peculiar nonmonotonicity is observed in the maximal values of each curve in e.g., Figs.~\ref{fig:phih-vs-M} and \ref{fig:kappa-vs-M}.}
\label{fig:M-dep}
\end{figure*}

The \emph{active}-$\rho$ branches exhibit some remarkable features. To showcase them, we plot as functions of the physical mass $M$ [obtained from Eq.~\eqref{eq:asymptotics_h}]: the value of the $\phi$ at the horizon in Fig~\ref{fig:phih-vs-M}, scalar charge $Q$ in Fig.~\ref{fig:Q-vs-M}, horizon area $A_h$ in Fig.~\ref{fig:Ah-vs-M}, and surface gravity $\kappa$ in Fig.~\ref{fig:kappa-vs-M}. These quantities show a common departure from sGB for lighter BHs in the direction \emph{away} from GR, and it is even stronger for the \emph{active}-$\rho$ branches, when they exist. This confirms and extends the expectations from the perturbative analysis. Some intuition for the origin of the branching can be gained from a test-field limit analysis of the effective $\rho$ equation \eqref{eq:effective_rho_eq}. The Gauss--Bonnet introduces a local (in space), attractive, inverse-square deformation to $V(\rho)$, affecting its extrema at scales $r \lesssim \ell_{\mathrm{NL}}$, i.e. an $r$-dependent effective potential $V_\text{eff}(\rho,r)$. Meanwhile, the scale $\mu^{-1}$ sets when modulus gradients become important. When both these effects are small, $\rho_0$ can only be asymptotically reached from its slightly displaced local equivalent, i.e. only the ordinary sGB-like branch exists with $\rho_h$ perturbatively close to $\rho_0$, well described by Eq.~\eqref{eq:rho_h-pert}. Instead, when both these effects are active, it becomes also possible to transition from one of the other displaced extrema of $V_\text{eff}(\rho,r_h)$, yielding up to two \emph{active}-$\rho$ solutions as found numerically above. For these to be good solutions not only for $\rho$, they must also reside outside the excluded region of Fig.~\ref{fig:rhoh-vs-rh}.

{ \textbf{\textit{Discussion}} }
sGB gravity is an EFT for a Goldstone boson. Natural UV-extensions of this theory involve gapped modes that become important at their respective mass scales $\mu$. This introduces phenomenologically interesting \emph{novel} behavior for BHs whose size approaches the scale of new physics $\mu^{-1}$.

As such UV-extension, we consider a complex scalar field with a symmetry-breaking potential. Our perturbative and numerical results indicate a stronger departure from GR compared to sGB in the low BH mass regime, when also the gapped mode's mass $\mu$ is of order $r_h^{-1}$. In particular, the scalar charge $Q$ can be significantly larger (Fig.~\ref{fig:Q-vs-M}). Observational consequences of this enhancement are expected for gravitational-wave physics, through: changes in the Quasi-Normal-Mode spectrum \cite{Konoplya:2011qq,Blazquez-Salcedo:2016enn,Bryant:2021xdh,Asuncion:2026efk}, stronger scalar-wave emission, stimulation of light scalar modes \cite{Barranco:2012qs} and superradiance for rotating BHs \cite{Hawking:1971tu,Detweiler:1980uk}, prompting to revisit the observational constraints on the sGB coupling $\alpha$ \cite{Lyu:2022gdr,Sanger:2024axs} from binary BH mergers. 

Of particular novelty is the emergence of multiple branches of solutions, each with a different, secondary scalar charge. Similar `hair' of discrete type has previously been associated only to non-shift-symmetric models with spontaneous scalarization \cite{Doneva:2017bvd}, suggesting a strong monopolar scalar radiation in the transition between branches. The latter can also happen because of accretion. The dynamical stability of the solutions is crucial for this analysis and shall be explored in the future. Beyond astrophysical scales, the branch structure may be particularly interesting in the mass distribution of low-mass primordial black holes (PBHs), see Refs.~\cite{Carr:2020xqk,Sasaki:2018dmp,Escriva:2022duf,Dokuchaev:2026kcp}). Naively, Hawking evaporation apparently drives BHs to the \emph{active}-$\rho$ regime, with their minimal possible mass larger than in sGB. 

The BH thermodynamics are also expected to be altered. While \emph{active}-$\rho$ solutions are hotter (following standard $T = \kappa/2\pi$ resulting from kinematics, see \cite{Kay:1988mu,Visser:2001kq,Deffayet:2011rh} and Fig.~\ref{fig:kappa-vs-M}) and have smaller area, suggesting a lower entropy, the heat capacity, $dM/dT$, is not always negative, Fig.~\ref{fig:Q-vs-M} and Eq.~\eqref{eq:charge}. This fundamental departure from the established GR (and sGB) behavior should be studied further and requires a more detailed study of entropy and boundary terms, see~\cite{Wald:1993nt,Iyer:1994ys,Liska:2023fdz,Hajian:2020dcq}.

In conclusion, the sGB model may find its immediate extension at a scale that is accessible by BHs during their evolution. At this threshold the phenomenological expectations are fundamentally altered and become sensitive to the details of the new physics.

{ \textbf{\textit{Acknowledgments}} }

The authors would like to thank E. Babichev for useful discussions. Numerical analysis has been partly carried out on the cluster Phoebe (CEICO, Prague). It is a great pleasure to thank Josef Dvo\v r\'a\v cek for the assistance with numerics. The work of G.T. was supported by IBS under the project code IBS-R018-D3. The work of L.G.T.\ was supported by Consejo Nacional de Investigaciones Cient\'ificas y T\'ecnicas (CONICET). The work of A.V.\ was supported by the Czech Science Foundation (GA\v{C}R) under Project No.\,24-13079S.

\bibliographystyle{utphys}
\bibliography{bibliography}

\appendix

\section{Field equations in spherical symmetry}

Consider the scalar field equations Eqs.~\eqref{EOM-rho}, \eqref{EOM-phi} and the metric equation associated to $T_{\mu\nu}$ given in Eq.~\eqref{Tmunu-CsGB}. We assume a spherically symmetric ansatz for the metric
\begin{equation}
    ds^2 = e^{A(r)} dt^2 - e^{B(r)} dr^2 - r^2 d\Omega^2,
\end{equation}
where we use the parametrization $h = e^A$ and $f = e^{-B}$, and for the two scalar fields
\begin{eqnarray}
    \rho &=& \rho(r), \\
    \phi &=& \phi(r).
\end{eqnarray}
With these ans\"atze, the $tt$ and $rr$ components of the metric field equations read, respectively
\begin{eqnarray}
    0 &=& 16\alpha \left( 1 - e^{-B} \right) \phi'' + 8 \alpha \left( 3 e^{-B} - 1 \right) B' \phi' + r^2 \rho'^2 \notag \\
    &&+ \frac{r^2 \rho^2 \phi'^2}{\rho_0^2} - 2r B' - 2 e^B \left[1 - r^2 V(\rho) \right] + 2, \qquad \label{Ett} \\
    0 &=& \left[ 1 - r^2 V(\rho) \right] \left( e^{B} \right)^2 + 12\alpha A' \phi' \notag \\
    &&- \Biggl( 1 + r A' + 4\alpha A' \phi' - \frac{r^2 \rho'^2}{2} - \frac{r^2 \rho^2 \phi'^2}{2\rho_0^2} \Biggr) e^B , \qquad \label{Err}
\end{eqnarray}
where, in order to allow for asymptotically flat solutions we assume $V(\rho_0) = 0$, with $\rho_0$ the vacuum solution in absence of gravity. The two angular components of the metric field equations are redundant. The $\rho$ and $\phi$ scalar equations are, respectively
\begin{eqnarray}
    0 &=& 2 r^2 \rho'' + r^2 \left( A' - B' \right) \rho' - \frac{2 r^2 \rho \phi'^2}{\rho_0^2}  + 4r \rho' \notag \\
    &&- 2r^2 V_\rho e^B, \label{Erho} \\
    0 &=& 8 \alpha \left( e^{-B} - 1 \right) A'' + \frac{2 r^2 \rho^2 \phi''}{\rho_0^2} + r^2 \left( A' - B' \right) \frac{\rho^2}{\rho_0^2} \phi' \notag \\
    &&+ \frac{4r^2 \rho \rho'}{\rho_0^2} \phi' + 4\alpha \left( 1 - 3 e^{-B} \right) A' B' \notag \\
    &&+ 4 \alpha \left( e^{-B} -1 \right) A'^2 + \frac{4r \rho^2}{\rho_0^2} \phi'. \label{Ephi}
\end{eqnarray}
We can solve the $rr$ equation \eqref{Err} algebraically for $e^B$ in terms of first derivatives of the other fields:
\begin{eqnarray}
    e^B = \frac{-\Gamma \pm \sqrt{\Gamma^2 - 48 \alpha A' \phi' \left(1 - r^2 V \right) }}{2\left(1 - r^2 V \right)},
\end{eqnarray}
where
\begin{eqnarray}
    \Gamma = - \left[ 1 + \left( r + 4 \alpha \phi' \right) A' - \frac{r^2 \rho'^2}{2} - \frac{r^2 \rho^2 \phi'^2}{2\rho_0^2} \right].
\end{eqnarray}
We must choose the $+$ branch so at infinity $e^B \to 1$ (provided everything decays sufficiently fast). Next, we can differentiate the $rr$ equation once w.r.t. $r$ to obtain an equation for $B'$
\begin{eqnarray}
    B' &=& \frac{1}{\left( 24 \alpha \phi' A' + \Gamma e^{B} \right)} \Bigg\{ \left[ 12 \alpha \phi' - e^B \left( r + 4 \alpha \phi' \right) \right] A'' \notag \\
    &&+ e^B r^2 \rho' \rho'' + \left[ 12 \alpha A' - e^B \left( 4 \alpha A' - \frac{r^2 \rho^2 \phi'}{\rho_0^2} \right) \right] \phi'' \notag \\
    &&+ e^B \left[ -A' + r \left( \rho'^2 + \frac{\rho^2}{\rho_0^2} \phi'^2 \right) + \frac{r^2 \rho \rho' \phi'^2}{\rho_0^2} \right] \notag \\
    &&- (2rV + r^2 V_\rho \rho') (e^B)^2 \Biggr\},
\end{eqnarray}
which, importantly, is linear in second derivatives, $A''$, $\rho''$ and $\varphi''$. This can be plugged back into the $tt$, $\rho$ and $\varphi$ equations, \eqref{Ett}, \eqref{Erho}, and \eqref{Ephi}, respectively. The resulting equations are also linear in second derivatives, composing a system of the form
\begin{eqnarray}
    \mathbb{M}(\rho, \rho', \phi')
    \begin{pmatrix}
     A'' \\
     \rho'' \\
     \phi''
    \end{pmatrix}
    = \mathbf{b}(\rho, \rho', \phi') ,
\end{eqnarray}
where $\mathbb{M}$ is a $3 \times 3$ matrix. Inverting this system allows to cast the equations as 
\begin{eqnarray} \label{inverted-eq-system}
    \begin{pmatrix}
     A'' \\
     \rho'' \\
     \phi''
    \end{pmatrix}
    =  \mathbb{M}^{-1}(\rho, \rho', \phi') \mathbf{b}(\rho, \rho', \phi') ,
\end{eqnarray}
amenable to numerical treatment.

\section{Regularity conditions}

The field equations in spherical symmetry \eqref{Ett}, \eqref{Err}, \eqref{Erho}, and \eqref{Ephi} must be supplemented by the condition of regular solutions at the horizon. Following the strategy of Ref.~\cite{Sotiriou:2014pfa}, we expand the equations in inverse powers of $A'$, which at the horizon behaves as $A' \to +\infty$, assuming the fields $\rho$ and $\phi$ and their derivatives remain finite. In what follows, the quantities in the right-hand sides are understood to be evaluated at the horizon.

First let us consider $e^B$, which we expect to go $e^B \to +\infty$. Indeed, provided
\begin{eqnarray} \label{eB_h-condition}
    r_h + 4 \alpha \phi'_h > 0,
\end{eqnarray}
then we get
\begin{eqnarray}
    e^B = \frac{\left(r + 4 \alpha \phi' \right)}{\left( 1 - r^2 V \right)} A' + \mathcal{O}(1).
\end{eqnarray}
Now we can proceed with the equations for $A''$, $\rho''$ and $\varphi''$ as they come out of the procedure described in Eq.~\eqref{inverted-eq-system}. At LO we get,
\begin{eqnarray}
    A'' &=& - \left[ 1 + \frac{4 \alpha}{D \left( 1 - r^2 V \right)} \left( P_2 \phi'^2 + P_1 \phi' + P_0 \right) \right] A'^2 \notag \\
    &&+ \mathcal{O}(A'), \\
    \rho'' &=& \frac{1}{D \left(1 - r^2 V \right)} \left[ \rho' N + \left(r + 4 \alpha \phi' \right) D \, V_\rho \right] A' \notag \\
    &&+ \mathcal{O}(1), \\
    \phi'' &=& - \frac{\left(r + 4 \alpha \phi' \right)}{D \left( 1 - r^2 V \right)} \left( P_2 \phi'^2 + P_1 \phi' + P_0 \right) A' \notag \\
    &&+ \mathcal{O}(1),
\end{eqnarray}
where
\begin{eqnarray} 
    D(\rho, \phi') &=& r^3 \rho^2 \left( r + 4 \alpha \phi' \right) \notag \\
    &&- \rho_0^2 \left(96 \alpha^2 + 64 \alpha^2 r^2 V - 128 \alpha^3 \phi' r V \right), \\
    N(\rho, \phi') &=& 48 \alpha^2 \rho_0^2 - r^4 \rho^2 - 8 \alpha \phi' r^3 \rho^2 \notag \\
    &&+ r^2 \left( r^4 \rho^2 V - 16 \alpha^2 \phi'^2 \rho^2 - 128 \alpha^2 \rho_0^2 V \right) \notag \\
    &&+ 16\alpha^2 V \left( \phi'^2 r^4 \rho^2 + 3 \rho_0^2 r^4 V - 32 \alpha^2 \rho_0^2 \phi'^2 \right)  \notag \\
    && + 8 \alpha \phi' r V \left( r^4 \rho^2 - 32 \alpha^2 \rho_0^2 \right) , \\
    P_0(\rho) &=&  4 \alpha \rho_0^2 \left( 3 - 2 r^2 V + r^4 V^2 \right), \\
    P_1(\rho) &=& r^3 \rho^2 - r^5 \rho^2 V + 96 r \alpha^2 \rho_0^2 V \notag \\
    &&- 32 \alpha^2 \rho_0^2 r^3 V^2 , \\
    P_2(\rho) &=& 4 \alpha \left( r^2 \rho^2 - r^4 \rho^2 V + 32 V \alpha^2 \rho_0^2 \right).
\end{eqnarray}

We now demand that both $\rho''$ and $\phi''$ remain finite at the horizon. In the near horizon expansions above both start at $\mathcal{O}(A')$, so we must demand the corresponding coefficients to vanish. First, notice that $\rho''$ at LO is linear in $\rho'$, so it is easy to fix $\rho'_h$ to make $\rho''_h$ finite:
\begin{eqnarray} \label{rhop_h}
    \rho_h'(\rho_h, \phi'_h) = - \frac{1}{N} \left(r + 4 \alpha \phi' \right) D \, V'.
\end{eqnarray}
Here $\rho'_h$ is a function of both $\rho_h$ and $\phi'_h$. Next we turn to $\phi''$. At LO the numerator contains the nonvanishing factor $r + 4 \alpha \phi'$ (see Eq.~\eqref{eB_h-condition}), and then another factor which is quadratic in $\phi'$, and independent on $\rho'$. Demanding the latter vanishes and solving for $\phi'_h$ gives
\begin{eqnarray}
    \phi'_h(\rho_h) = \frac{- P_1 + \sqrt{P_1^2 - 4 P_2 P_0}}{2P_2}.
\end{eqnarray}
The choice of $+$ branch is to recover the known sGB solutions in the proper limit. In order for $\phi'_h$ to be real, 
\begin{eqnarray}
    \frac{P_1^2 - 4 P_2 P_0}{\left(1 - r_h^2 V \right)^2} = r_h^6 \rho_0^4 \mathcal{D}_h &=& r_h^6 \rho_h^4 - 192 \alpha^2 \rho_0^2 r_h^2 \rho_h^2 \notag \\
    &&+ 128 \alpha^2 \rho_0^2 V \left( r_h^4 \rho_h^2 - 48 \alpha^2 \rho_0^2 \right) \notag \\
    &&+ 1024 \alpha^4 \rho_0^4 r_h^2 V^2 > 0.
\end{eqnarray}
This is condition \eqref{existence} from the main text. Interestingly, replacing any branch of $\phi'_h$ into the equation for $A''$ at LO, it identically gives
\begin{eqnarray}
    A'' = -A'^2,
\end{eqnarray}
which is just the statement that $e^A$ has a linear zero at the horizon.

\section{Perturbative solutions}

Following Ref.~\cite{Sotiriou:2014pfa} we start by considering the ans\"atze,
\begin{eqnarray}
    e^{A(r)} &=& \left(1 - \frac{2m}{r} \right) \left( 1 + \sum_{n=1}^\infty A_n(r) \tilde{\alpha}^n \right)^2 \, , \\
    e^{B(r)} &=& \left(1 - \frac{2m}{r} \right)^{-1} \left( 1 + \sum_{n=1}^\infty B_n(r) \tilde{\alpha}^n \right)^2 \, , \\
    \phi'(r) &=& \sum_{n=1}^\infty \phi'_n(r) \tilde{\alpha}^n \, , \\
    \rho(r) &=& \rho_0 + \sum_{n=1}^\infty \rho_n(r) \tilde{\alpha}^n \, .
\end{eqnarray}
where here $m = r_h/2$ at all orders in $\tilde{\alpha} = \alpha/r_h^2$. Replacing these expressions into the field equations \eqref{Ett}, \eqref{Err}, \eqref{Erho}, and \eqref{Ephi}, and solve iteratively in $\tilde{\alpha}$. For the potential $V(r)$ we consider \eqref{potential}. By power counting one can already see that $A(r)$, $B(r)$, and $\rho(r)$ only contain even powers of $\tilde{\alpha}$, while $\phi(r)$ only contains odd powers of $\tilde{\alpha}$.

In what follows we present the order by order solutions which satisfy the correct boundary conditions, i.e. that tend to zero at $r \to \infty$, and are regular at $r = 2m$. This can always be achieved by uniquely fixing the integration constants. 

For $n=1$ we get,
\begin{eqnarray}
    A_1(r) &=& B_1(r) = \rho_1(r) = 0 \, \\
    \phi'_1(r) &=& -\frac{8m}{r^4} \left( r^2 + 2mr + 4m^2 \right) \, ,
\end{eqnarray}
while for $n=2$ we can easily get $A_2(r)$, $B_2(r)$, and $\phi_2(r)$
\begin{widetext}
\begin{eqnarray}
    A_2(r) &=& \frac{2 \, m^{3} r^{2}}{15 \, {\left(2 \, m - r\right)}^{2}} {\left(\frac{6400 \, m^{5}}{r^{8}} - \frac{4736 \, m^{4}}{r^{7}} - \frac{288 \, m^{3}}{r^{6}} - \frac{1552 \, m^{2}}{r^{5}} + \frac{960 \, m}{r^{4}} + \frac{40}{r^{3}} + \frac{294}{m r^{2}} - \frac{147}{m^{2} r}\right)} \, , \\
    B_2(r) &=& -\frac{2 \, m^{4}}{15 \, {\left(2 \, m - r\right)}} {\left(\frac{147}{m^{3}} + \frac{14720 \, m^{3}}{r^{6}} - \frac{384 \, m^{2}}{r^{5}} - \frac{240 \, m}{r^{4}} - \frac{2080}{r^{3}} - \frac{120}{m r^{2}} - \frac{120}{m^{2} r}\right)} \, \\
    \phi'_2(r) &=& 0 \, ,
\end{eqnarray}
\end{widetext}
but we already face a difficulty for $\rho_2(r)$. The equation for the latter being
\begin{eqnarray}
    0 &=& {\left(2 m r^{2} - r^{3}\right)} \rho_2'' + 2 {\left(m r - r^{2}\right)} \rho_2' + 2 \mu^{2} r^{3} \rho_2 \notag \\
    &&- \frac{64 \, {\left(2 m r^{2} - r^{3}\right)} {\left(4 m^{2} + 2 m r + r^{2}\right)}^{2} m^{2}}{r^{8} \rho_{0}} \, . \notag \\
\end{eqnarray}
This equation includes kinetic, potential (proportional to $\mu^2$), and source terms.  In order to find a closed-form solution we must consider some approximation. On the one hand, when $\mu m \gg 1$, the potential term dominates over the kinetic term and the solutions to approach sGB. On the other hand, in the $\mu m \ll 1$ the regime the solutions are expected to significantly differ from sGB. 

\subsection{Regime $\mu m \ll 1$ ($ \mu^{-1} \gg r_h $)}
We find a solution,
\begin{eqnarray}
    \rho_2(r) &=& \frac{8m}{225 \, r^{6} \rho_{0}} (800 \, m^{5} + 1056 \, m^{4} r + 1110 \, m^{3} r^{2} \notag \\
    &&\qquad\qquad+ 340 \, m^{2} r^{3} - 195 \, m r^{4} - 1095 \, r^{5}) \notag \\  
    &&+ \mathcal{O}\left( \frac{m^2\mu^2}{\rho_0} \right) \, . 
\end{eqnarray}
Notice that at this order, this is the only place where we see a departure from sGB, as characterized by the $\rho_0^{-1}$ factor, which it is still not showing up in the metric functions $A(r)$ and $B(r)$, nor in $\phi(r)$, all of which still have the same form as in sGB. At higher orders, $\rho_2(r)$ acts as a source and will propagate this difference to the rest of the fields. We need to go up to $n=4$ to specifically see the backreaction on the metric functions. From now on we will proceed with under the approximation $\mu \to 0$, which restricts the validity of the results. In this context, we may still recover the sGB limit by taking $\rho_0 \to \infty$, which in the above solution gives, as expected, $\rho_2(r) \to 0$.

\begin{widetext}
At $n=3$, we again only have a nontrivial solution for $\phi'_3(r)$,
\begin{eqnarray}
    A_3(r) &=& B_3(r) = \rho_3(r) = 0 \, \\
    \phi'_3(r) &=& -\frac{16 \, m}{225 \, r^{10} \rho_{0}^{2}} \Bigl(2035200 \, m^{8} \rho_{0}^{2} + 668160 \, m^{7} r \rho_{0}^{2} + 168000 \, m^{6} r^{2} \rho_{0}^{2} - 157440 \, m^{5} r^{3} \rho_{0}^{2} \notag \\
    &&- 28560 \, m^{4} r^{4} \rho_{0}^{2} + 7920 \, m^{3} r^{5} \rho_{0}^{2} + 15780 \, m^{2} r^{6} \rho_{0}^{2} + 4380 \, m r^{7} \rho_{0}^{2} - 15 \, r^{8} \rho_{0}^{2} - 25600 \, m^{8} \notag \\
    &&- 46592 \, m^{7} r - 58816 \, m^{6} r^{2} - 37088 \, m^{5} r^{3} - 8080 \, m^{4} r^{4} + 35440 \, m^{3} r^{5} + 19080 \, m^{2} r^{6} \notag \\
    &&+ 8760 \, m r^{7}\Bigr) + \mathcal{O}(\mu^2) \, . 
\end{eqnarray}
Notice the presence of terms dependent on $\rho_0$, signaling a departure from sGB, which then vanish for $\rho_0 \to \infty$.

Finally, for $n=4$ we obtain
\begin{eqnarray}
    A_4(r) &=& \frac{1}{779625 \, r^{12} \rho_{0}^{2}} \Bigl[ 769950720000 \, m^{12} \rho_{0}^{2} - 5203968000 \, m^{12} \notag \\
    &&- {\left(330684930 \, m \rho_{0}^{2} + 421203727 \, m\right)} r^{11} - 2 \, {\left(405560115 \, m^{2} \rho_{0}^{2} + 421203727 \, m^{2}\right)} r^{10} \notag \\
    &&- 4 \, {\left(460338300 \, m^{3} \rho_{0}^{2} + 396583747 \, m^{3}\right)} r^{9} - 8 \, {\left(241727985 \, m^{4} \rho_{0}^{2} + 371963767 \, m^{4}\right)} r^{8} \notag \\
    &&- 80 \, {\left(13083678 \, m^{5} \rho_{0}^{2} - 7743547 \, m^{5}\right)} r^{7} + 32 \, {\left(136054035 \, m^{6} \rho_{0}^{2} + 168184919 \, m^{6}\right)} r^{6} \notag \\
    &&- 64 \, {\left(428848380 \, m^{7} \rho_{0}^{2} - 231112619 \, m^{7}\right)} r^{5} - 896 \, {\left(76405185 \, m^{8} \rho_{0}^{2} - 3038747 \, m^{8}\right)} r^{4} \notag \\
    &&- 258048 \, {\left(696695 \, m^{9} \rho_{0}^{2} + 18307 \, m^{9}\right)} r^{3} + 774144 \, {\left(37815 \, m^{10} \rho_{0}^{2} - 14258 \, m^{10}\right)} r^{2} \notag \\
    &&+ 12902400 \, {\left(16730 \, m^{11} \rho_{0}^{2} - 701 \, m^{11}\right)} r \Bigr] + \mathcal{O}(\mu^2) \, , \\
    B_4(r) &=& -\frac{1}{779625 \, {\left(2 \, m - r\right)}} \Bigl[ 330684930 \, m + \frac{421203727 \, m}{\rho_{0}^{2}} \notag \\
    &&+ \frac{2}{r^{12} \rho_{0}^{2}} \Bigl( 
     5128510464000 \, m^{13} \rho_{0}^{2} - 43997184000 \, m^{13} + 3465 \, {\left(65307 \, m^{2} \rho_{0}^{2} - 42632 \, m^{2}\right)} r^{11} \notag \\
     &&- 6930 \, {\left(5493 \, m^{3} \rho_{0}^{2} + 21316 \, m^{3}\right)} r^{10} - 4620 \, {\left(501819 \, m^{4} \rho_{0}^{2} + 42632 \, m^{4}\right)} r^{9} \notag \\
     &&- 27720 \, {\left(190413 \, m^{5} \rho_{0}^{2} + 290978 \, m^{5}\right)} r^{8} - 11088 \, {\left(743055 \, m^{6} \rho_{0}^{2} + 202472 \, m^{6}\right)} r^{7} \notag \\
     &&+ 36960 \, {\left(1619319 \, m^{7} \rho_{0}^{2} + 23948 \, m^{7}\right)} r^{6} + 31680 \, {\left(1252371 \, m^{8} \rho_{0}^{2} + 2133656 \, m^{8}\right)} r^{5} \notag \\
     &&+ 443520 \, {\left(206463 \, m^{9} \rho_{0}^{2} + 43483 \, m^{9}\right)} r^{4} - 788480 \, {\left(1056411 \, m^{10} \rho_{0}^{2} + 14018 \, m^{10}\right)} r^{3} \notag \\
     &&- 8515584 \, {\left(35795 \, m^{11} \rho_{0}^{2} + 6963 \, m^{11}\right)} r^{2} - 3870720 \, {\left(170455 \, m^{12} \rho_{0}^{2} + 14776 \, m^{12}\right)} r    
    \Bigr) \Bigr] \notag \\
    &&+ \mathcal{O}(\mu^2) \, , \\
    \phi'_4(r) &=& 0 \, , \\
    \rho_4(r) &=& \frac{4}{900466875 \, r^{12} \rho_{0}^{3}} \Bigl[ 12567582720000 \, m^{12} \rho_{0}^{2} - 136604160000 \, m^{12} \notag \\
    &&- 3465 \, {\left(28105540 \, m \rho_{0}^{2} + 67271073 \, m\right)} r^{11} - 3465 \, {\left(78237160 \, m^{2} \rho_{0}^{2} + 67271073 \, m^{2}\right)} r^{10} \notag \\
    &&- 4620 \, {\left(56851180 \, m^{3} \rho_{0}^{2} + 6564273 \, m^{3}\right)} r^{9} - 6930 \, {\left(12550000 \, m^{4} \rho_{0}^{2} - 37510527 \, m^{4}\right)} r^{8} \notag \\
    &&+ 11088 \, {\left(61457780 \, m^{5} \rho_{0}^{2} + 68210427 \, m^{5}\right)} r^{7} + 18480 \, {\left(61841240 \, m^{6} \rho_{0}^{2} + 41127987 \, m^{6}\right)} r^{6} \notag \\
    &&+ 31680 \, {\left(17632460 \, m^{7} \rho_{0}^{2} + 13736007 \, m^{7}\right)} r^{5} - 388080 \, {\left(4279360 \, m^{8} \rho_{0}^{2} + 1063719 \, m^{8}\right)} r^{4} \notag \\
    &&- 2759680 \, {\left(684400 \, m^{9} \rho_{0}^{2} + 248481 \, m^{9}\right)} r^{3} + 14902272 \, {\left(365280 \, m^{10} \rho_{0}^{2} - 43997 \, m^{10}\right)} r^{2} \notag \\
    &&+ 270950400 \, {\left(38090 \, m^{11} \rho_{0}^{2} - 1367 \, m^{11}\right)} r \Bigr] + \mathcal{O}(\mu^2) \, .
\end{eqnarray}
\end{widetext}
At this order we also observe $\rho_0$ terms that vanish when $\rho_0 \to \infty$, containing corrections to the pure sGB solutions.

At this stage we may compute physical quantities of interest. First, the value of the modulus at the horizon,
\begin{eqnarray} \label{rho_h_of_m}
    \rho_h &=& \rho_{0} - \frac{3511 \, {\alpha}^{2}}{3600 \, m^{4} \rho_{0}} - \frac{10796992439 \, {\alpha}^{4}}{5762988000 \, m^{8} \rho_{0}} \notag \\
    &&- \frac{226241560627 \, {\alpha}^{4}}{102453120000 \, m^{8} \rho_{0}^{3}}\,,
\end{eqnarray}
which satisfies $\rho_h < \rho_0$ in this regime, confirming a departure from sGB.
The physical black-hole mass $M$ (ADM mass) can be obtained from the asymptotic behavior for $r \to \infty$ of $B(r)$. Up to order $\alpha^4$ it reads,
\begin{eqnarray} \label{Mofm}
    M &=& m + \frac{49 \, {\alpha}^{2}}{40 \, m^{3}} + \frac{408253 \, {\alpha}^{4}}{246400 \, m^{7}} \notag \\
    &&+ \frac{60171961 \, {\alpha}^{4}}{28512000 \, m^{7} \rho_{0}^{2}} \, .
\end{eqnarray}
This can now be used to, perturbatively, replace $m$ in favor of $M$ where required For example for $\rho_h$,
\begin{eqnarray} \label{rho_h_of_M}
\rho_h &=& \rho_{0} - \frac{3511 \, {\alpha}^{2}}{3600 \, M^{4} \rho_{0}} - \frac{9584377919 \, {\alpha}^{4}}{1440747000 \, M^{8} \rho_{0}} \notag \\
&&- \frac{226241560627 \, {\alpha}^{4}}{102453120000 \, M^{8} \rho_{0}^{3}} \, .
\end{eqnarray}
Similarly, from the asymptotic behavior of $\phi(r)$ we may read the scalar charge $Q$,
\begin{eqnarray}
    Q = \frac{2 \, {\alpha}}{M} + \frac{73 \, {\alpha}^{3}}{30 \, M^{5}} \, ,
\end{eqnarray}
which we are already expressing in terms of $M$. 

It is also interesting to compute quantities at the black-hole horizon, such as the area,
\begin{eqnarray}
    A_h &=& 16 \, \pi M^{2} - \frac{196 \, \pi {\alpha}^{2}}{5 \, M^{2}} - \frac{666319 \, \pi {\alpha}^{4}}{3850 \, M^{6}} \notag \\
    &&- \frac{60171961 \, \pi {\alpha}^{4}}{891000 \, M^{6} \rho_{0}^{2}} \, ,
\end{eqnarray}
surface gravity,
\begin{eqnarray}
    \kappa &=& \frac{1}{4 \, M} + \frac{73 \, {\alpha}^{2}}{240 \, M^{5}} + \frac{4010597 \, {\alpha}^{4}}{2217600 \, M^{9}} \notag \\
    &&+ \frac{22423691 \, {\alpha}^{4}}{44352000 \, M^{9} \rho_{0}^{2}} \, ,
\end{eqnarray}
which can be verified to satisfy, up to the current level of accuracy, the exact relation $Q = 8\alpha \kappa$, as derived from the regularity of Eq.~\eqref{phase-eq-int-once}. We also compute the value of phase field
\begin{eqnarray}
    \phi_h = \frac{11 \, {\alpha}}{6 \, M^{2}} + \frac{4298 \, {\alpha}^{3}}{675 \, M^{6}} + \frac{111017 \, {\alpha}^{3}}{45360 \, M^{6} \rho_{0}^{2}} \, .
\end{eqnarray}
We may finally consider the inverse relation to Eq.~\eqref{Mofm},
\begin{eqnarray}
    m &=& M - \frac{49 \, {\alpha}^{2}}{40 \, M^{3}} - \frac{303503 \, {\alpha}^{4}}{49280 \, M^{7}} \notag\\
    &&- \frac{60171961 \, {\alpha}^{4}}{28512000 \, M^{7} \rho_{0}^{2}} \,,
\end{eqnarray}
which gives the size of the black hole, $r_h = 2m$, in terms of the GR value $2M$, for the same mass $M$. 

To finally obtain the expressions as in the main text, Eqs.~\eqref{eq:rho_h-pert}, \eqref{eq:M_perturbative}, \eqref{eq:A_h_perturbative}, \eqref{eq:phi_h_perturbative}, and \eqref{eq:kappa_perturbative}, we simply note that in the current regime we have $\rho_0 \ll 1$ and just keep the leading departure from sGB. For $\rho_h$ this happens at $\mathcal{O}(\alpha^2)$, while for the remaining quantities we must up to $\mathcal{O}(\alpha^4)$ where generically there is a pure sGB contribution going as $\mathcal{O}((\rho_0)^0)$, and one originating from the complex scalar behavior at $\mathcal{O}(\rho_0^{-2})$. Therefore, in this regime the latter dominates and it is the one we keep at $\mathcal{O}(\alpha^4)$. The last step is to identify $\ell_\rho^4 = \frac{\alpha^2}{\rho_0^2}$.

\subsection{Regime $\mu m \gg 1$ ($ \mu^{-1} \ll r_h $)}

For completeness we also consider the limit of large $\mu m$. The solution for $\rho_2(r)$ after dropping the derivative terms is
\begin{eqnarray}
    \rho_2\left(r\right) &=& \frac{64 \, {\left(2 \, m r^{2} - r^{3}\right)} {\left(4 \, m^{2} + 2 \, m r + r^{2}\right)}^{2} m^{2}}{2 \, \mu^{2} r^{11} \rho_{0}} \notag \\
    &&+ \mathcal{O}\left( \frac{1}{m^4 \mu^4 \rho_0} \right) \, .
\end{eqnarray}
We can iterate in the same way as before and compute the nonzero $\phi_3(r)$, $A_4(r)$, $B_4(r)$, and $\rho_4(r)$. Then it is possible to compute the physical quantities. We obtain, for the asymptotic quantities
\begin{eqnarray}
    M &=& m + \frac{49 \, {\alpha}^{2}}{40 \, m^{3}} + \frac{408253 \, {\alpha}^{4}}{246400 \, m^{7}} \notag \\
    &&+ \frac{130309 \, {\alpha}^{4}}{1441440 \, m^{9} \mu^{2} \rho_{0}^{2}} \, , \\
    Q &=& \frac{2 \, {\alpha}}{M} + \frac{73 \, {\alpha}^{3}}{30 \, M^{5}} \, ,
\end{eqnarray}
while for the horizon quantities
\begin{eqnarray}
    \phi_h &=& \frac{11 \, {\alpha}}{6 \, M^{2}} + \frac{4298 \, {\alpha}^{3}}{675 \, M^{6}} + \frac{11383 \, {\alpha}^{3}}{110880 \, M^{8} \mu^{2} \rho_{0}^{2}} \, , \\
    \rho_h &=& \rho_0 \, , \\
    \kappa &=& \frac{1}{4 \, M} + \frac{73 \, {\alpha}^{2}}{240 \, M^{5}} + \frac{4010597 \, {\alpha}^{4}}{2217600 \, M^{9}} \notag \\
    &&+ \frac{589 \, {\alpha}^{4}}{24024 \, M^{11} \mu^{2} \rho_{0}^{2}} \,.
\end{eqnarray}

\section{Numerical method}

We studied numerical solutions to the full system of equations in spherical symmetry. Throughout we kept the theory parameters $\alpha=0.015$ and $\lambda=0.25$ fixed. The effective equation for $u = \rho/\rho_0$, Eq.~\eqref{eq:effective_rho_eq}, is analogous to a mechanical equation for the 1-dimensional $u(t)$ (identifying $t = r$) motion of a particle of mass $\mu^{-2}$, subject to a time-dependent potential
\begin{eqnarray}
    U(u, t) = \frac{1}{f(t)} \left(- \frac{8 \ell_\text{NL}^6}{u^2} S(t) - V(u) \right) \, ,    
\end{eqnarray}
and dissipation coefficient
\begin{eqnarray}
    \Gamma(t) = \mu^{-2} \left( \frac{2}{t} + \frac{h'}{2h} + \frac{f'}{2f} \right)    
\end{eqnarray}
A solution that approaches the vacuum at infinity, $u \to 1$, corresponds in this analogy to siting on the top of the inverted potential $-V(u)$ located at $u=1$, at infinite time $t$. 

To find such solution, we applied a shooting procedure by choosing the \emph{initial condition}, i.e. the value $0 < u_h \leq 1$ satisfiying Eq.~\eqref{existence}, at a position $\epsilon = 10^{-5}$ from the horizon $r = r_h$. The rest of the horizon data is fixed algebraically as in Eq.~\eqref{eq:horizon-data}. A generic solution either overshoots or undershoots $u = 1$, and the transition between these two behaviors as $u_h$ is varied signals the existence of a solution with the right asymptotics for some intermediate value. The panels in Fig.~\ref{fig:r-dep} show the solutions found in this way for $\mu = \{ 0.04, 0.05, 0.06, 0.1 \}$, for fixed $r_h = 1.4$.

For Figs.~\ref{fig:rhoh-vs-rh} and \ref{fig:M-dep}, $\mu = \{ 0.03, 0.05, 0.1, 0.2, 0.3 \}$, while $r_h$ was varied in the order of $10^3$ discrete steps in the range $ 0.4559 \lesssim r_h \leq 2.4$, where the lower bound is the minimum allowed size in pure sGB [$(192\alpha^2)^{1/4}$], for $\alpha=0.015$.

\end{document}